# The Unified Logging Infrastructure for Data Analytics at Twitter

George Lee, Jimmy Lin, Chuang Liu, Andrew Lorek, and Dmitriy Ryaboy

Twitter, Inc.
@GeorgeJLee @lintool @chuangl4 @mrtall @squarecog

## ABSTRACT

In recent years, there has been a substantial amount of work on large-scale data analytics using Hadoop-based platforms running on large clusters of commodity machines. A less-explored topic is how those data, dominated by application logs, are collected and structured to begin with. In this paper, we present Twitter's production logging infrastructure and its evolution from application-specific logging to a unified "client events" log format, where messages are captured in common, well-formatted, flexible Thrift messages. Since most analytics tasks consider the user session as the basic unit of analysis, we pre-materialize "session sequences", which are compact summaries that can answer a large class of common queries quickly. The development of this infrastructure has streamlined log collection and data analysis, thereby improving our ability to rapidly experiment and iterate on various aspects of the service.

## 1. INTRODUCTION

The advent of scalable, distributed, and fault-tolerant frameworks for processing large amounts of data—especially the Hadoop open-source implementation of MapReduce [7]—has made it easier for organizations to perform massive data analytics to better understand and serve customers and users. These ideas are, of course, not new: business intelligence, OLAP, and data warehouses have been around for decades, but we argue that the field has recently entered a new phase. The rise of social media and user-generated content, shift to cheap commodity clusters, and a growing ecosystem of open-source tools create new challenges as well as rich opportunities for organizations in search of deeper insights from their vast stores of accumulated data.

Twitter has built a Hadoop-based platform for large-scale data analytics running Pig [26, 10] on a cluster of several hundred machines; see [19] for an earlier description. It serves more "traditional" business intelligence tasks such as cubing to support "roll up" and "drill down" of multi-dimensional data (for example, feeding dashboards and other visualizations) as well as more advanced predictive analytics such as machine learning [18]. These capabilities, however, are not the subject of this paper. Instead, we focus on how log data, on the order of a hundred terabytes uncompressed in aggregate each day, arrive at the data warehouse from tens of thousands of production servers, are structured in a consistent yet flexible format to support data analytics, and how compressed digests called "session sequences" are materialized to speed up certain classes of queries. The overall goal is to streamline our ability to experiment and iterate in a data-driven fashion based on analysis of log data.

We begin with an overview of our message delivery infrastructure built on top of Scribe, an existing piece of open-source software. We then explain the motivation for developing a unified logging framework where all events are captured in common, well-formatted, semi-structured Thrift messages. Events are identified using a consistent, hierarchical naming scheme, making it easy to identify data associated with a particular Twitter client, page, tab, or even UI widget. These client event logs simplify analyses greatly, but at the cost of more verbose logs than might be necessary for any individual application, which results in increased overall log volume and thus lower query performance. The logs are arranged in partial chronological order when they arrive in the data warehouse, but the most common unit of analysis is the user session—therefore, analytical queries usually begin with scans over potentially terabytes of data, followed by a group-by to reconstruct the user sessions; only then does the actual analysis begin. To support this common case operation, we pre-materialize compressed summaries of client events organized by user sessions. These are about fifty times smaller than the original logs, but support large classes of common queries. We call these "session sequences", and they address many of the performance issues with client events logs.

The goal of this paper is to share with the community our experiences in developing Twitter's unified logging infrastructure. There has been a lot of published research studies [1, 25, 4, 21] and industry experiences [28, 5] at the intersection of databases and Hadoop-based data analytics, but we note a paucity of details on the interactions between producers of this large-scale data, the infrastructure that manages them, the engineers who build analytical pipelines, the data scientists who mine the vast data stores for insights, and the final consumers of the analytical results. We aim to fill this gap, and view this work as having three contributions: First, we briefly describe the messaging infrastructure at Twitter that enables robust, scalable delivery and aggre-gation of log data totaling around one hundred terabytes per day. Second, we discuss the evolution from application-specific logging to a unified logging format that greatly simplifies downstream analyses. Third, we describe a technique for pre-materializing summaries of user sessions that can be viewed as a common-case optimization for large classes of information requests.

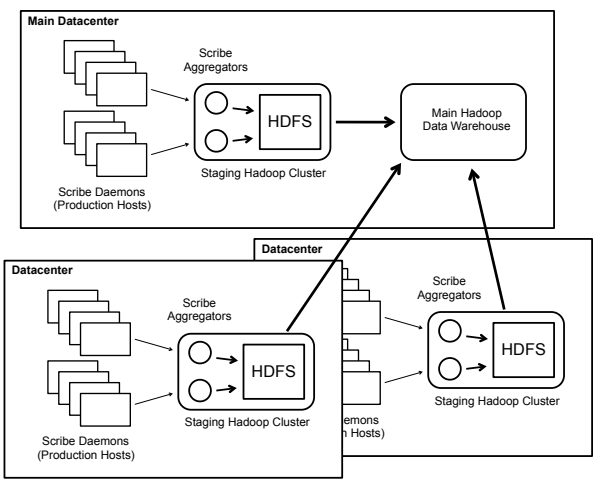

Figure 1: Illustration of Twitter's Scribe infrastructure. Scribe daemons on production hosts send log messages to Scribe aggregators, which deposit aggregated log data onto per-datacenter staging Hadoop clusters. Periodic processes then copy data from these staging clusters into our main Hadoop data warehouse.

We readily admit that, with perhaps the exception of some small "tricks", nothing presented in this paper is novel from a research perspective. However, neither is our architecture common knowledge within the community and best practices adopted by all. We are not aware of similar expositions of "war stories" on what it's like to wrestle with terabytes of logs on a daily basis in a continuously evolving and growing production environment. Therefore, we believe that our experiences are valuable to the community in contributing to a growing body of knowledge on large-scale data analytics.

## 2. SCRIBE INFRASTRUCTURE

Scribe is a system for aggregating high volumes of streaming log data in a robust, fault-tolerant, distributed manner. Its development began at Facebook in 2007, and in October 2008 Scribe was open sourced. Although it has since been augmented by other systems, Scribe remains an integral part of Facebook's logging infrastructure. A description of how it fits into the overall scheme of Facebook's data collection pipeline can be found in [29]. More recent open-source systems such as Apache Flume[1] and LinkedIn's Kafka [17] offer similar functionality.

Twitter's Scribe infrastructure is illustrated in Figure 1, and is similar to the design presented in [29]. A Scribe daemon runs on every production host and is responsible for sending local log data across the network to a cluster of dedicated aggregators in the same datacenter. Each log entry consists of two strings, a category and a message. The category is associated with configuration metadata that determine, among other things, where the data is written.

In our implementation, Scribe daemons discover the hostnames of the aggregators through ZooKeeper [12], a robust, open-source, wait-free coordination system. ZooKeeper provides to clients the abstraction of a set of data nodes (called znodes), organized in a hierarchical namespace, much like a filesystem. Aggregators register themselves at a fixed location using what is known as an "ephemeral" znode, which exists only for the duration of a client session; the Scribe daemons consult this location to find a live aggregator they can connect to. If an aggregator crashes or is restarted by an administrator, the ZooKeeper session is terminated and the ephemeral znode disappears. Scribe daemons simply check Zookeeper again to find another live aggregator. The same mechanism is used for balancing load across aggregators.

The aggregators in each datacenter are co-located with a staging Hadoop cluster. Their task is to merge per-category streams from all the server daemons and write the merged results to HDFS (of the staging Hadoop cluster), compressing data on the fly. Another process is responsible for moving these logs from the per-datacenter staging clusters into the main Hadoop data warehouse. It applies certain sanity checks and transformations, such as merging many small files into a few big ones and building any necessary indexes. Lastly, it ensures that by the time logs are made available in the main data warehouse, all datacenters that produce a given log category have transferred their logs. Once all of this is done, the log mover pipeline atomically slides an hour's worth of logs into the main data warehouse. The entire pipeline is robust with respect to transient failures—Scribe daemons discover alternative aggregators via ZooKeeper upon aggregator failure, and aggregators buffer data on local disk in case of HDFS outages. We do not delve into too much detail here, as these are fairly mature systems with widely-known best practices.

At the end of the log mover pipeline, logs arrive in the main data warehouse and are deposited in per-category, per-hour directories (e.g., `/logs/category/YYYY/MM/DD/HH/`). Within each directory, log messages are bundled in a small number of large files. The ordering of messages within each file is unspecified, although in practice they are partially time-ordered (since each file contains multiple files from aggregators that have been rolled up). This has practical consequences for downstream analysis, as we describe later.

## 3. TOWARD UNIFIED LOGGING

This section describes our experiences in moving from application-specific logging to a unified log format. We motivate this evolution by highlighting some of the challenges encountered when trying to actually perform analyses over the data in search of insights. As it is impossible to frame this discussion without referring to the analytics platform itself, we provide a brief description here, but refer the reader to previously-published details [19, 18].

When the contents of records are well-defined, they are serialized using one of two serialization frameworks. Protocol Buffers[2] (protobufs) and Thrift[3] are two language-neutral data interchange formats that provide compact encoding of structured data. Both support nested structures, which allow developers to capture and describe rich data in a flexible way. In addition, both protobufs and Thrift are extensible, allowing messages to gradually evolve over time while preserving backwards compatibility. For example, messages can be augmented with additional fields in a completely transparent way. For logging, there is a preference for Thrift since Thrift is also used widely in Twitter as a language-agnostic RPC framework and thus it is familiar to developers across the organization.

---

[1] http://incubator.apache.org/flume/
[2] http://code.google.com/p/protobuf/
[3] http://thrift.apache.org/

In a Hadoop job, different record types produce different types of input key-value pairs for the mappers, each of which requires custom code for deserializing and parsing. Since this code is both regular and repetitive, it is straightforward to use the serialization framework to specify the data schema, from which the serialization compiler generates code to read, write, and manipulate the data. This is handled by our system called Elephant Bird,[4] which automatically generates Hadoop record readers and writers for arbitrary Protocol Buffer and Thrift messages.

Although the analytics platform at Twitter is built around Hadoop, actual production jobs and *ad hoc* queries are performed mostly using Pig, a high-level dataflow language that compiles into physical plans that are executed on Hadoop [26, 10]. Pig (via a language called Pig Latin) provides concise primitives for expressing common operations such as projection, selection, group, join, etc. This conciseness comes at low cost: Pig scripts approach the performance of programs directly written in Hadoop Java. Yet, the full expressiveness of Java is retained through a library of custom UDFs that expose core Twitter libraries (e.g., for tokenizing and manipulating tweets).

Production analytics jobs are coordinated by our workflow manager called Oink, which schedules recurring jobs at fixed intervals (e.g., hourly, daily). Oink handles dataflow dependencies between jobs; for example, if job $B$ requires data generated by job $A$, then Oink will schedule $A$, verify that $A$ has successfully completed, and then schedule job $B$ (all while making a best-effort attempt to respect periodicity constraints). Finally, Oink preserves execution traces for audit purposes: when a job began, how long it lasted, whether it completed successfully, etc. One common Oink data dependency is the log mover pipeline, so once logs arrive in the main data warehouse, dependent jobs are automatically triggered. Each day, Oink schedules hundreds of Pig scripts, which translate into tens of thousands of Hadoop jobs.

## 3.1 Motivation

Initially, all applications, and in some cases, even parts of applications, defined their own, custom structure that they logged via Scribe. This approach afforded significant flexibility and allowed for very fast application logging development, as developers had to come up with a simple logging object definition in Thrift, start using it, and supply the necessary metadata to link their logs to the Thrift object description. Elephant Bird automatically handled Hadoop-specific code generation for processing Thrift messages using either Hadoop MapReduce or Pig. However, we soon discovered that this approach also had several drawbacks when it came to using and managing the data. Here, we provide a sampling of the challenges.

---

[4]http://github.com/kevinweil/elephant-bird

Consider, for example, automation of common summarization tasks ("how many events of each type did we see yesterday?"). This required custom, per-category configuration. Application developers adopt different approaches for logging common primitives such as timestamps; programmers accustomed to different language environments have conflicting conventions for naming fields and variables. This resulted in code generating and analyzing logs being littered with CamelCase, smallCamelCase, as well as snake_case (and occasionally, even the dreaded camel_Snake). As a specific example: Is the user id field uid, userId, userid, or user_id (of course, not ruling out user_Id)? Other mundane issues such as whether delimited characters were tabs, spaces, or a particular control character, as well as the handling of embedded delimiters often caused problems—the wrong setting in a Pig script, for example, would yield no output or complete garbage. Although most of these issues can rightly be characterized as "annoyances", accumulation of small minor annoyances impedes rapid development.

In our setup, each application writes logs using its own Scribe category. For simplicity, we implemented a one-to-one mapping between the Scribe category and the HDFS directory in the main data warehouse where the imported logs ultimately reside (/logs/category/). In practice, this creates a resource discovery problem, in that there are literally several dozen Scribe categories, many non-intuitively named or whose contents have substantially diverged from when the Scribe category was first established, making the category name meaningless at best or misleading at worst. Since the application developers are often disjoint from the data scientists who perform the analyses downstream, it is sometimes difficult, particularly for new hires, to even figure out what logs are available. As documentation falls out of sync with code over time, we rely increasingly on tacit knowledge passed along through group mailing lists and by word of mouth.

Finding the logs themselves (i.e., "resource discovery") is merely the first challenge. With application-specific logging, log messages are often difficult to understand due to the myriad of different formats. Some logs are completely regular Thrift. Some are a union of regular formats, in that log messages can take one of a finite number of different formats, where each format is regular. Some are in "natural language", where certain phrases serve as the delimiters. Some are semi-structured, with repeating and optional fields. Finally, the "worst" are log messages that themselves have dynamic, rich internal structure. An example is frontend logs, which capture rich user interactions (tweet impressions, link click-throughs, etc.) in JSON format. These JSON structures are often nested several layers deep to capture all relevant parameters of the user interaction. At analysis time, it is often difficult to make sense of the logs: for example, in the JSON structure, what fields are obligatory, what fields are optional? For each field, what is the type and range of values? Can it be null, and if so, what does that indicate? Even obtaining a complete catalog of all possible message types is difficult. Internal documentation is almost always out of date, and the knowledge lives primarily in the team of developers who created the applications (who themselves were often fuzzy on the details of code they had written months ago). To get around this issue, engineers on the analytics team often had to read frontend code to figure out the peculiarities of a message of interest, or induce the message format manually by writing Pig jobs that scraped large numbers of messages to produce key-value histograms. Needless to say, both of these alternatives are slow and error-prone.

| Component | Description | Example |
|---|---|---|
| client | client application | `web`, `iphone`, `android` |
| page | page or functional grouping | `home`, `profile`, `who_to_follow` |
| section | tab or stream on a page | `home`, `mentions`, `retweets`, `searches`, `suggestions` |
| component | component, object, or objects | `search_box`, `tweet` |
| element | UI element within the component | `button`, `avatar` |
| action | actual user or application action | `impression`, `click`, `hover` |

Table 1: Hierarchical decomposition of client event names.

A large part of this particular issue stems from JSON's lack of a fixed schema. This affords the developer great flexibility in describing complex data with nested structures, but it also increases the potential for abuse in creating overly complex messages without regard to downstream analytical needs. JSON-specific issues aside, we are still left with the fact that application-specific logging leads to a proliferation of different formats, making it difficult for data scientists to understand all the data necessary for their jobs. Of course, we would like to make their lives easier without burdening application developers upstream, for example, by forcing them into overly-rigid structures. Like many design problems, striking a good balance between expressivity and simplicity is difficult.

Beyond the myriad of logging formats, challenges are compounded when attempting to join data from multiple sources, as most non-trivial analytics tasks require. Assuming that one has already found the possibly inconsistently-named fields on which to join and divined the format of each separate application log, making sense of joined data sometimes required Herculean effort. The issue is that most analyses are based on a notion of user sessions, from which we can infer casual relationships: for example, the user entered a search query, browsed results, and then clicked on a link. There was no consistent way across all applications to easily reconstruct the session, except based on timestamps and the user id (assuming they were actually logged). So, Pig analysis scripts typically involved joins (by user id), group-by operations, followed by ordering with respect to timestamps and other *ad hoc* bits of code to deal with application-specific idiosyncrasies (glossing over the fact that timestamps and other supposedly common elements were captured in half a dozen different ways). This process was slow and error prone, since even minor inconsistencies often led to garbage or empty output.

As a response to the challenges described above, the "client events" unified logging format was developed. We turn our attention to this next.

## 3.2 Client Events

"Client events" is a relatively recent effort within Twitter to develop a unified logging framework that simplifies analyses without imposing substantial additional burden on application developers.

Generalizing the notion of Scribe categories, we imposed a hierarchical six-level naming scheme for *all* events (comprised of client, page, section, component, element, action), outlined in Table 1. This six-level decomposition aligns with the view hierarchy of Twitter clients. For example, in the case of the main web client (i.e., the twitter.com site), the namespace corresponds to the page's DOM structure, making it possible to automatically generate event names and thereby enforce consistent naming. This makes it possible to perform a reverse mapping also; that is, given only the event name, we can easily figure out based on the DOM where that event was triggered.

As a specific example, the following event

`web:home:mentions:stream:avatar:profile_click`

is triggered whenever there is an image profile click on the avatar of a tweet in the mentions timeline for a user on twitter.com ("reading" the event name from right to left). To combat the dreaded `camel_Snake`, we imposed consistent, lowercased naming. This hierarchical namespace makes it easy to slice-and-dice categories of events with simple regular expressions to focus on an *ad hoc* grouping of interest. For example, analyses could be conducted on all actions on the user's home mentions timeline on twitter.com by considering `web:home:mentions:*`; or track profile clicks across all clients (twitter.com, iPhone, Android, etc.) with `*:profile_click`.

Furthermore, Oink jobs automatically aggregate counts of events according to the following schemas:

```
(client, page, section, component, element, action)
(client, page, section, component, *, action)
(client, page, section, *, *, action)
(client, page, *, *, *, action)
(client, *, *, *, *, action)
```

These counts are presented as top-level metrics in our internal dashboard, further broken down by country and logged in/logged out status. Thus, without any additional intervention from the application developer, rudimentary statistics are computed and made available on a daily basis.

As an alternative design, we had considered a looser tree-based model for naming client events, i.e., the event namespace could be arbitrarily deep. The advantage is that it avoids empty component names (e.g., if a page doesn't have multiple sections, the section component is simply empty) and supports details as fine-grained as necessary. Flexibility, however, comes at the cost of complexity and the fact that the top-level aggregates above would be more difficult to automatically compute. Ultimately, we decided against this design and believe that we made the correct decision.

A client event itself is a Thrift structure that contains the components shown in Table 2 (slightly simplified, but still preserving the essence of our design). The `event_initiator` specifies whether the event was triggered on the client side or the server side, and whether the event was user initiated or application initiated (for example, a user's timeline polls for new tweets automatically without user intervention). Note that *all* client events contain fields for user id, session id (based on browser cookie or other similar identifier), and IP address. This was an explicit attempt to address the issues discussed in Section 3.1—where inconsistent tracking of these elements made joining data across disparate sources difficult. Since every client event has these fields, with exactly the same semantics, a simple group-by suffices to accurately reconstruct user sessions (of course, timestamps are still important for ordering events). By extension, standardizing the location and names of these fields allows us to implement consistent policies for log anonymization.

| Field | Description |
| --- | --- |
| `event_initiator` | {client, server} × {user, app} |
| `event_name` | event name |
| `user_id` | user id |
| `session_id` | session id |
| `ip` | user's IP address |
| `timestamp` | timestamp |
| `event_details` | event details |

Table 2: Definition of a client event.

Finally, the `event_details` field holds event-specific details as key-value pairs. For example, in the profile click event described above, the details field would hold the id of the profile clicked on. For an event corresponding to a search result, the `event_details` field would hold the target URL, rank in the result list, and other such information. Since different teams can populate these key-value pairs as they see fit, the message structure can be flexibly extended without any central coordination.

It is worth emphasizing that this logging format is imposed across all Twitter clients: not only the twitter.com web site, but also clients on the iPhone, iPad, Android phones, etc. To the extent possible, events of the same type across different clients are given the same name. This is made possible by the consistent design language imposed in the latest iteration of Twitter clients. For example, all clients have a section for viewing a user's mentions (other tweets that reference the user); an impression means the same thing, whether on the web client or the iPhone. The practical effect of this is that Pig scripts written to analyze behavior on one client can be ported over to another client with relative ease.

In summary, client events form the foundation of our unified logging infrastructure in two senses of the word "unified": first, in that all log messages share a common format with clear semantics, and second, in that log messages are stored in a single place (as opposed to different Scribe category silos with application-specific logging). This leads to a number of advantages, especially compared to application-specific logging:

- A consistent, hierarchical event namespace across all Twitter clients and a common message format makes logs easier to understand and more approachable. This also enables automatic materialization of top-level metrics.

- Common semantics for different fields in the Thrift messages makes it easier to reconstruct user session activity.

- A single location for all client event messages simplifies the resource discovery issue (knowing where to find what logs). This also obviates the need for joins for many common types of analyses.

The cost of the above advantages is that clients must implement this specification (whereas before developers were free to log however they wished). Overall, we feel this is a worthwhile tradeoff, since the additional burden imposed on the application developer is minimal.

## 4. SESSION SEQUENCES

The evolution toward unified logging enables additional manipulation of log data to support large classes of common queries. In this section, we describe pre-materialized digests of user sessions called "session sequences".

### 4.1 Motivation

Despite the advantages of a unified logging format, there is one downside: the logs tend to be more verbose, even after compression. Unfortunately, we were unable to accurately quantify the space tradeoff since application-specific logging was converted gradually to the client events format over a period of several months, during which Twitter traffic increased measurably. Thus, it is difficult to attribute the source of log growth.

With the unified client event logs, analyses in terms of user sessions became simpler—the developer no longer needed to join multiple disparate data sources, and the presence of consistent user ids and session ids allowed sessions to be reconstructed with greater accuracy. However, reconstructing sessions remained processing intensive—essentially, a large group-by across potentially terabytes of data. These jobs routinely spawned tens of thousands of mappers and clogged our Hadoop jobtracker, performing large amounts of brute force scans and data shuffling, all before useful analysis actually began. Furthermore, since analyses tended to focus on fresh (i.e., newly imported) logs, we observed frequent contention for a relatively small set of HDFS blocks. We wished to address this performance issue.

We discovered that large classes of queries could be answered using only the sequence of clients event names within a user session—the hierarchical namespace provides a lot of information alone. Critically, these queries did not require "peering" into the details of the Thrift messages. Examples are queries that involve computing click-through rate (CTR) and follow-through rate (FTR) for various features in the service: how often are search results, who-to-follow suggestions, trends, etc. clicked on within a session, with respect to the number of impressions recorded? Similarly, what fraction of these events led to new followers? In these cases, it suffices to know that an impression was followed by a click or follow event—it is not necessary to know *what* URL was being clicked on or *who* exactly was being followed. Of course, deeper analyses might require breaking down clicks and follows in terms of topical categories or users' interests, etc., but the coarse-grained CTR/FTR calculations remain a common-case query. Note that top-level CTR/FTR metrics can be pre-computed and displayed in dashboards, but data scientists often desire statistics for arbitrary subsets of users (e.g., causal users in the U.K. who are interested in sports), which require *ad hoc* queries.

Another common class of queries that require only event names involves navigation behavior analysis, which focuses on how users navigate within Twitter clients. Examples questions include: How often do users take advantage of the "discovery" features? How often do tweet detail expansions lead to detailed profile views? Answers to these questions provide important feedback for refining site and client app design. Given that the event names directly reflect the organization of the view hierarchy of the clients, the names alone are sufficient to answer these questions.

In Pig, scripts for computing these types of statistics would operate over the client event logs. The first operation is usually to project onto the event name (discarding all other fields), retaining only those of interest, followed by a group-by to reconstruct the session. The early projection and filtering keeps the amount of data shuffling (from mappers to reducers) to a reasonable amount, but the large numbers of brute force disk scans remain a substantial bottleneck.

Since most of our Pig scripts begin by reconstructing user sessions, it made sense to simply pre-materialize the sessions. However, instead of materializing the entire client event Thrift structure, we generate compact, compressed summaries of the sequence of client event *names*, thus addressing the inefficiencies associated with the common scan-and-project operations (we discuss this design decision in more detail below). These are the session sequences.

## 4.2 Implementation

A session sequence is defined as a sequence of symbols $S = \{s_0, s_1, s_2...s_n\}$ where each symbol is drawn from a finite alphabet $\Sigma$. We define a bijective mapping between $\Sigma$ and the universe of event names, so that the sequence of symbols corresponds to the sequence of client events that comprise the user session. Each symbol is represented by a unicode code point, such that any session sequence is a valid unicode string, i.e., sequence of unicode characters. Furthermore, we define the mapping between events and unicode code points (i.e., the dictionary) such that more frequent events are assigned smaller code points. This in essence captures a form of variable-length coding, as smaller unicode points require fewer bytes to physically represent. For example, the event:

```
web:home:mentions:stream:avatar:profile_click
```

might get mapped to \u0235. Unicode comprises 1.1 million available code points, and it is unlikely that the cardinality of our alphabet will exceed this. Since each session sequence is a valid unicode string, we can express analyses in terms of regular expressions and other common string manipulations. Note that these session sequences are not meant for direct human consumption, and we provide tools for accessing and manipulating them (more below).

Construction of session sequences proceeds in two steps. Once all logs for one day have been successfully imported into our main data warehouse, Oink triggers a job that scans the client event logs to compute a histogram of event counts. These counts, as well as samples of each event type, are stored in a known location in HDFS (this is used by our automatically-generated catalog described in Section 4.3). The histogram construction job also builds a client event dictionary that maps the event names to unicode code points, based on frequency as noted above.

In a second pass, sessions are reconstructed from the raw client event logs. This is accomplished via a group-by on `user_id` and `session_id`; following standard practices, we use a 30-minute inactivity interval to delimit user sessions. These sequences of event names are then encoded using the dictionary. The following relation is materialized on HDFS (slightly simplified):

```
user_id: long, session_id: string, ip: string,
session_sequence: string, duration: int
```

To summarize, a session sequence is simply a unicode string that captures the names of the client events that comprise the session in a compact manner. Alongside this representation, we store the user and session ids, the IP address associated with the session, as well as the session duration in seconds (i.e., temporal interval between the first and last event in the session). Note that other than the overall session duration, session sequences do not preserve any temporal information about the events (other than relative ordering). This was an explicit design choice to support compact encoding, but we lose the ability to discriminate temporal gaps between successive events.

It is worthwhile to discuss our choice to materialize these session sequences, and not to simply alter the physical layout of the raw client event Thrift messages. Raw client event logs are cumbersome to work with for two independent reasons: first, their large size translates into a lot of brute force disk scans, and second, the physical on-disk organization requires large group-by operations to reconstruct user sessions before useful analysis can actually begin. We had originally considered an alternative design where we simply reorganized (i.e., rewrote) the *complete* Thrift messages by reconstructing user sessions. This would have solved the second issue (large group-by operations) but would have little impact on the first (too many brute force scans). To mitigate that issue, we could adopt a columnar storage format such as RCFile [11]. However, this solution primarily focuses on reducing the running time of each map task; without modification, RCFiles would not reduce the number of mappers that are spawned for large analytics jobs (which can easily number in the tens of thousands) and the associated job-tracker traffic. Hadoop tasks have relatively high startup costs, and we would like to avoid this overhead as well. Our materialized session sequences have the advantage in that they are about fifty times smaller than the original client event logs and therefore they address both the group-by and brute force scan issues at the same time.

In summary, queries over session sequences are substantially faster than queries over the raw client event logs, both in terms of lower latency and higher throughput. The cost of increased performance is that session sequences are restricted to answering only a certain class of queries. Nevertheless, as we describe in Section 5, they support a wide range of applications.

## 4.3 Client Event Catalog

From our experiences with application-specific logging, we learned that reducing the barrier to entry for comprehending log formats not only makes data scientists more productive, but also helps foster a data-driven culture across the company by providing easy access to people who wouldn't otherwise perform analytics. Part of this is maintaining concise, up-to-date documentation, which our automatically-generated client event catalog assists with.

Client events represent a vast improvement in comprehensibility over application-specific logging, by virtue of two features: First, the uniform log format means that one can focus on the semantics of the log message and not the format. Second, the consistent, hierarchical namespace and the correspondence between the client view hierarchy and event names means that client events are to a large extent self-documenting.

To further enhance the accessibility of our client event logs, we have written an automatically-generated event catalog and browsing interface which is coupled to the daily job of building the client event dictionary. The interface lets users browse and search through the client events in a variety of ways: hierarchically, by each of the namespace components, and using regular expressions. For each event, the interface provides a few illustrative examples of the complete Thrift structure (via the sampling process while building the client event dictionary, described above). Finally, the interface allows developers to manually attach descriptions to the event types.

Since the event catalog is rebuilt every day, it is always up to date. Even if most of the entries do not contain developer-supplied descriptions, the catalog remains immensely useful as a single point of entry for understanding log contents. The only remaining issue, a holdover from the application-specific logging days, is that without additional documentation, in some cases it is difficult to fully understand the semantics of event_details with sample messages alone. For example: Which keys are always present? Which are optional? What are the ranges for values of each key? In principle, it may be possible to infer from the raw logs themselves, but we have not implemented this functionality yet. On balance, though, we find that supporting extensible key-value pairs in event_details is a reasonable compromise between not imposing too much burden on the developer (e.g., forcing a rigid schema) and simplifying analyses downstream (e.g., by moving away from arbitrary JSON).

## 5. APPLICATIONS

The client event logs and session sequences form the basis of a variety of applications, a few of which we describe in this section.

### 5.1 Summary Statistics

Due to their compact size, statistics about sessions are easy to compute from the session sequences. A series of daily jobs generate summary statistics, which feed into our analytical dashboard called BirdBrain. The dashboard displays the number of user sessions daily and plotted as a function of time, which when coupled with a variety of other metrics, lets us monitor the growth of the service over time and spot trends. We also provide the ability to drill down by client type (i.e., twitter.com site, iPhone, Android, etc.) and by (bucketed) session duration.

### 5.2 Event Counting

As previously described, session sequences greatly speed up simple queries that involve *ad hoc* counting of events. A typical Pig script might take the following form (slightly simplified):

```
define CountClientEvents
  CountClientEvents('$EVENTS');
raw = load '/session_sequences/$DATE/'
  using SessionSequencesLoader();
...
generated = foreach raw generate
  CountClientEvents(symbols);
grouped = group generated all;
count = foreach grouped generate SUM(generated);
dump count;
```

We begin by specifying the $EVENTS we wish to count in the initialization of the CountClientEvents user-defined function; here, an arbitrary regular expression can be supplied which is automatically expanded to include all matching events (via the dictionary that provides the event name to unicode code point mapping). The variable $DATE specifies the date over which to execute the query. A custom Pig loader abstracts over details of the physical layout of session sequences, transparently parsing each field in the tuple and handling decompression.

Typically, what happens next is a series of transformations that are query-specific: for example, if the data scientist wishes to restrict consideration of the user population by various demographics criteria, a join with the users table followed by selection with the appropriate criteria would ensue. Finally, the CountClientEvents UDF is executed, which returns the number of client events found in the particular session sequence. Since a session sequence is simply a unicode string, the UDF translates into string manipulations after consulting the client event dictionary for the event name to unicode code point mapping. A Pig group all followed by SUM tallies up the counts.

A common variant of the above script is a replacement of SUM by COUNT. Instead of returning the total event count, this returns the number of user sessions that contain at least one instance of a particular client event. These analyses are useful for understanding what fraction of users take advantage of a particular feature.

### 5.3 Funnel Analytics

Beyond simple counting, one class of complex queries that the session sequences support is broadly known as "funnel analytics" [2, 22]. Originally developed in the context of e-commerce sites, these analyses focus on user attention in a multi-step process. A classic example is the purchase checkout flow; a canonical funnel might be: users visit their shopping cart, enter billing and shipping addresses, select shipping options, enter payment details, and finally confirm. The funnel analogy is particularly apt because it captures abandonment at each step, e.g., users select a shipping option but never enter payment details. In general, an e-commerce site wants to maximize the number of users that flow through the entire funnel (i.e., complete a purchase) along with related metrics (e.g., total revenue). The number and nature of each step in the funnel plays an important role; for example, shortening the number of steps potentially comes at the cost of making each step in the flow more complex; on the other hand, users have little patience for activities that require too many steps. In addition, the layout and design of each step can have a substantial impact on the abandonment rate. Companies typically run A/B tests to optimize the flow [16], for example, varying the page layout of a particular step or number of overall steps to assess the impact on end-to-end metrics (e.g., revenue or number of successfully-completed purchases).

In the context of Twitter, there are a variety of complex funnels. An important one is the signup flow, which is the sequence of steps taken by a user to join the service. The flow needs to be easy for new users but difficult for spammers attempting to create fraudulent accounts. It needs to show novice users how to use the service without being boring. Part of the signup flow includes presenting new users with lists of suggested accounts to follow, broken down into interest categories, the contents of which impact the user experience. The complexity of the signup flow admits many design choices, both in the length of the funnel and the complexity of each step. Given that a vibrant user base is the lifeblood of a social media service, growing an engaged audience is one of the core priorities of the organization.

While backend analytics tools cannot actually help in the design process, they are absolutely critical in assessing the effectiveness of each design. To help, we have created a UDF for defining funnels:

```
define Funnel ClientEventsFunnel('$EVENT1'
  '$EVENT2', ...);
```

where the data scientist specifies an arbitrary number of stages that make up the funnel in terms of client event names. After evaluation of the UDF and grouping similar to the simple event counting script, the output might be something like:

```
(0, 490123)
(1, 297071)
...
```

which tells us how many of the examined sessions entered the funnel, completed the first stage, etc. This particular UDF translates the funnel into a regular expression match over the session sequence string. Once again, since session sequences are simply unicode strings, such code is easy to write. In fact, it is in principle possible to expose the session sequences themselves to manipulation using arbitrary regular expressions (mediated by a thin wrapper that handles the dictionary translation), although we have not yet found a need for such functionality.

Building on the above example, other variants are easy. Translating these figures into the number of users (as opposed to sessions) is simply a matter of applying the unique operator in Pig prior to summing up the per-stage counts. From here, additional statistics such as per-stage abandonment can be easily computed.

## 5.4 User Modeling

Whereas funnel analytics aims to answer specific questions about user behavior, we have a parallel branch of work on the more open-ended question of identifying "interesting" user behavior. Since session sequences are simply symbol sequences drawn from a finite alphabet, we can borrow techniques derived from natural language processing (NLP).

A useful tool commonly-used in NLP is language modeling [23, 14]. Language models define a probability distribution over sequences of symbols, $P(w_1 w_2 w_3 \ldots w_{n-1} w_n) \equiv P(w_1^n)$. Due to the extremely large number of parameters involved in estimating such a model, it is customary to make the *Markov assumption*, that the sequence histories only depends on prior local context. That is, an *n*-gram language model is equivalent to a *(n-1)*-order Markov model. Thus, we can approximate $P(w_k|w_1^{k-1})$ as follows:

$$\begin{aligned} \text{bigrams:} \quad & P(w_k|w_1^{k-1}) & \approx P(w_k|w_{k-1}) \\ \text{trigrams:} \quad & P(w_k|w_1^{k-1}) & \approx P(w_k|w_{k-1}w_{k-2}) \\ \text{$n$-grams:} \quad & P(w_k|w_1^{k-1}) & \approx P(w_k|w_{k-n+1}^{k-1}) \end{aligned}$$

Metrics such as cross entropy and perplexity [23, 14] can be used to quantify how well a particular *n*-gram model "explains" the data, which gives us a sense of much "temporal signal" there is in user behavior [20, 24]. Intuitively, how the user behaves right now is strongly influenced by immediately preceding actions; less so by an action 5 steps ago, and even less by an action 15 steps ago. Language modeling techniques allow us to more precisely quantify this.

Another useful concept borrowed from the NLP community is the notion of collocations, or commonly-occurring patterns of words [6, 9]; see Pearce [27] for a survey. A simple example is "hot dog", where the two terms "hot" and "dog" co-occur much more frequently than one would expect if the words simply appeared independently. Collocations frequently have non-compositional meanings, as in this case. Applying the analogy to session sequences, it is possible to extract "activity collocates" [20], which represent potentially interesting patterns of user activity. We have begun to perform these types of analyses, borrowing standard techniques from text processing such as pointwise mutual information [6] and log-likelihood ratios [9].

## 6. ONGOING WORK

The client events unified logging infrastructure went into production in Spring 2011. Transitioning application-specific logs over to client event logs happened gradually (although the logs were almost immediately useful). The session sequences went into production Winter 2011. We are presently working on a number of extensions, some of which are described below.

The first attempts to address the limitation of session sequences, in that they capture only the event names, and not details important for certain types of analyses. To complement session sequences, we have recently deployed into production a generic indexing infrastructure for handling highly-selective queries called Elephant Twin; this was first described in [19] and recently presented at the 2012 Hadoop Summit.[5] The infrastructure is general, although client event logs represent one of the first applications.

Although it is fairly obvious that indexes are important for query performance, there are a number of design and operational challenges that render Hadoop integration a non-trivial task (see our previous discussion [19]). While it is true that indexing solutions have previously been proposed and implemented in the Hadoop context, our approach is fundamentally different. Indexes are available in Hive, but the feature is integrated higher up in the stack and so indexes only benefit Hive queries. Another approach, dubbed "Trojan layouts" embeds indexes in HDFS block headers, the consequence of which is that indexing requires rewriting data [8] (cf. [13]). This unfortunately is not practical with petabytes of data and in an environment where the indexing scheme might evolve over time. Furthermore, Trojan layouts introduce significant complexity since they alter HDFS replication behavior and make fundamental modifications to the filesystem that hamper scalability.

Our Elephant Twin indexing framework integrates with Hadoop at the level of InputFormats, which means that applications and frameworks higher up the Hadoop stack can transparently take advantage of indexes "for free". In Pig, for example, we can easily support push-down of select operations. Our indexes reside *alongside* the data (in contrast to Trojan layouts), and therefore re-indexing large amounts of data is feasible. For example, we perform full-text indexing of all tweets for our internal tools; as our text processing libraries improve (e.g., better tokenization), we drop all indexes and rebuild from scratch; in fact, this has already happened several times during the past year.

---

[5]http://engineering.twitter.com/2012/06/twitter-at-hadoop-summit.html

Finally, we are attempting to move beyond well-known techniques such as iteration via A/B testing into more exploratory approaches, even beyond those outlined in Section 5.4. The field of natural language processing provides fertile intellectual ground from which we can draw rather straightforward analogies to user behavior: application of language models and collocation extraction are two simple examples. More advanced (but speculative) techniques include applying automatic grammar induction techniques to learn hierarchical decompositions of user activity [15]. For example, we might learn that many sessions break down into smaller units that exhibit a great deal of cohesion (each with rich internal structure), in the same way that a simple English sentence decomposes into a noun phrase and a verb phrase. Somewhat surprisingly, there is a rich interplay between techniques for natural language processing and analysis of biological sequence (e.g., DNA). Bridging these two worlds, we can take inspiration from biological sequence alignment [3] to answer questions like: "What users exhibit similar behavioral patterns?" This type of "query-by-example" mechanism would help in understanding what makes Twitter user engaged and "successful", and what causes one to abandon the service.

In the realm of exploratory user modeling, but unrelated to both text processing and biological sequence analysis, we are also using advanced visualization techniques [30] to provide data scientists a *visual* interface for exploring sessions—the hope is that interesting behavioral patterns will map into distinct visual patterns, such that insights will literally "leap off the screen".

In summary, all these ongoing activities are enabled by unified client events logging and session sequences. We have only begun to scratch the surface in terms of possibilities, and are excited by the solid foundation that the infrastructure provides.

## 7. CONCLUSIONS

Data is the lifeblood of many organizations today, but it's sometimes easy to forget that accumulating vast data warehouses is pointless in and of itself. What's important are the insights that the data reveal about users and customers—and obtaining these insights in a timely fashion. Supporting these activities requires several different components: a robust message delivery mechanism, flexibility at the point where log messages are generated, and ease with which those messages can be manipulated downstream to perform data analytics. For the first, Twitter uses Scribe and Thrift, both of which are mature and widely-deployed. For the second and third points, our first-generation application-specific architecture made logging easy, but analysis difficult. We believe that the unified client events logging infrastructure strikes a better balance between the desires of application developers (flexible logging) and the needs of data scientists (easy data analysis). Session sequences are a way to make common case queries even easier and faster. This, and other ongoing projects comprise our efforts to streamline the end-to-end analytics lifecycle, from messages logged at production servers to insights delivered by data scientists.

## 8. ACKNOWLEDGMENTS

The logging infrastructure described in this paper is part of Twitter's analytics stack, to which numerous engineers from the analytics team and from other parts of the organization have contributed. This work would not have been possible without advice from our data scientists, who provide valuable guidance on what functionalities are actually useful in their quest for insights. We'd like to thank Gilad Mishne and Eric Sammer for useful comments on earlier drafts of this paper.